# Robustness of the thermal Hall effect close to half-quantization in a field-induced spin liquid state

J.A.N. Bruin[1*], R.R. Claus[1], Y. Matsumoto[1], N. Kurita[2], H. Tanaka[2], H. Takagi[1,3*]

[1] Max Planck Institute for Solid State Research, 70569 Stuttgart, Germany

[2] Department of Physics, Tokyo Institute of Technology, Tokyo 152-8551, Japan

[3] Department of Physics, The University of Tokyo, Bunkyo, Tokyo 113-0033, Japan

**Thermal signatures of fractionalized excitations are a fingerprint of quantum spin liquids (QSLs)[1]. In the $J_{eff}$=1/2 honeycomb magnet α-RuCl$_3$, a QSL state emerges upon applying an in-plane magnetic field $H_{//}$ greater than the critical field $H_{C2}$ ≈ 7 T along the *a*-axis, where the thermal Hall conductivity ($k_{XY}/T$) was reported to take on the half-quantized value $k_{HQ}/T$. This finding was discussed as a signature of an emergent Majorana edge mode predicted for the Kitaev QSL[2]. The $H_{//}$- and *T*-range of the half-quantized signal[2-4] and its relevance to a Majorana edge mode are, however, still under debate[5,6]. Here we present a comprehensive study of $k_{XY}/T$ in α-RuCl$_3$ with $H_{//}$ up to 13 T and *T* down to 250 mK, which reveals the presence of an extended region of the phase diagram with $k_{XY}/T$ ≈ $k_{HQ}/T$ above $H_{C2}$, in particular across a plateau-like "plane" for $H_{//}$ > 10 T and *T* < 6.5 K. From 7 T up to ~10 T, $k_{XY}/T$ is suppressed to zero upon cooling to lowest temperature without any plateau-like behavior and exhibits correlations with complex anomalies in the longitudinal thermal conductivity ($k_{XX}$) and magnetization around 10 T. The results are in support of a topological state with a half-quantized $k_{XY}/T$ and suggest an interplay with crossovers or weak phase transitions beyond $H_{C2}$ in RuCl$_3$.**

A prime candidate for experimentally accessible QSLs is the Kitaev QSL, which has an exactly solvable ground state with itinerant and localized Majorana fermions[7]. Whilst these are charge-neutral, the itinerant Majorana fermions carry heat and are therefore expected to contribute to thermal transport[8]. Applying a magnetic field gaps the bulk Majorana bands, leading to a topologically protected chiral edge current[9]. This edge state carries a thermal Hall conductivity per layer divided by temperature of $\kappa_{XY}^{2D}/T = \frac{\pi^2 k_B^2}{6h}$, one half of that of an equivalent electronic edge state in the Quantum Hall effect due to the fractionalized nature of the Majorana fermion.

The search for material candidates of the Kitaev model has focused on Mott insulators[10] with $5d^4$ $Ir^{4+}$ and $4d^4$ $Ru^{3+}$ with strong spin-orbit coupling on a honeycomb lattice[11-13]. α-RuCl$_3$ is a prime candidate which shows antiferromagnetic zig-zag order below $T_N \approx 7.5$ K. This magnetic order is suppressed with an in-plane magnetic field of $H_{C2} \approx 7$ T[2,14-21], where the magnetic moment is not yet fully saturated, revealing a region of the phase diagram where a QSL is established. A thermal Hall effect with a magnitude close to the half quantized value $k_{HQ}/T$ (corresponding to $\kappa_{XY}^{2D}/T$ per atomic plane) was reported in the QSL region with $H_{||}$ in the $a$-axis (perpendicular to the Ru-Ru bond direction)[2-4], which was discussed as a signature of the Majorana edge state expected for the Kitaev quantum spin liquid.

To establish the presence of edge state, however, the robustness of the half quantized thermal hall effect should be demonstrated over a reasonably wide range of magnetic fields and temperatures. Several studies report a plateau-like region in field around 5 K with a magnitude close to $\kappa_{HQ}/T$[2-4]. These studies, however, are limited down to ~3.5 K and the plateau behavior as a function of $T$ is not as clear as it is in field. The plateau onset fields differ between studies, ranging from $H_{||} = 7.8$ T[2] to 9.9 T[3]. In preparing this manuscript, we noticed that Czajka and coworkers[22] reported a broad $k_{XY}/T$ dome without clear plateau-like behavior around $\kappa_{HQ}/T$. Recently, new anomalies in the magnetocaloric effect, specific heat and magnetic Grüneisen parameter were reported around $H_{\parallel}$ = 10 T[23-25] in the QSL region (> $H_{C2}$), suggesting that the phase diagram may be even more complex. The question of the relationship between these anomalies and the thermal Hall signature may be of fundamental importance to understanding the nature of the field induced QSL, and more specifically the possible topological edge state.

Here, we present comprehensive measurements of $k_{XY}$, $k_{XX}$ and the magnetic susceptibility (d$M$/d$H$) in a $T$-range from 150 mK to 9 K and $H_{\parallel}$ up to 13 T ($H_{\parallel}$ along $a$-axis) on an α-RuCl$_3$ single crystal, which reveals that $k_{XY}/T$ stays close to $k_{HQ}/T$ over a reasonably wide range of both temperature and magnetic field, below $T$ = 6.5 K and from $H_{\parallel}$ ~10 T up to at least 13 T, in favor of the presence of the protected half-quantized plateau. Below ~10 T, $k_{XY}/T$ is suppressed to zero upon lowering temperature below 6.5 K, which appears to correlate with the multiple high-field ($H_{\parallel}$ > $H_{C2}$) anomalies present in both $k_{XX}$ and d$M$/d$H$.

**Magnetic-field-induced anomalies in the thermal conductivity $k_{XX}$**

The temperature dependence of $k_{XX}$ reproduces previous reports[2] very well and displays a sharp minimum at the antiferromagnetic ordering temperature $T_N$ = 7.5 K, which moves to lower temperature as $H_{\parallel}$ increases and eventually fades out above $H_{C2}$ = 7.1 T, tracing the extent of the antiferromagnetically ordered region (Figs. 1a,e). As a function of $H_{\parallel}$, the magnitude of $k_{XX}$ first decreases to a broad minimum around 7 T and then rapidly increases (Fig. 1b). The data below 2 K reveal additional features in $k_{XX}(H_{\parallel})$: the broad minimum splits into two at $H_{C1}$ = 6.05 T and $H_{C2}$ = 7.1 T. These fields coincide with two sharp peaks in d$M$/d$H$ (Fig. 1c), which correspond to the two reported transitions from a zigzag-ordered phase I to an intermediate ordered phase II, and from phase II to the high-field paramagnetic phase, respectively[23-28]. The suppression of $k_{xx}$ at the magnetic phase transitions is consistent with the interpretation that $k_{xx}$ is dominated by phonons which scatter off magnetic excitations[29].

At higher in-plane magnetic fields and below 5 K, further anomalies can be seen in $k_{xx}$: a broad maximum at $H_P$~10 T flanked by broad minima around $H_{D1}$~8.8 T and $H_{D2}$~10.7 T (Fig. 1b). Features corresponding to $H_{D1}$ and $H_P$ can be clearly identified as a broad shoulder and a sudden drop in d$M$/d$H$ (Fig. 1c). We argue that these high-field anomalies in $k_{xx}$ arise from crossovers or very weak phase transitions.

In analogy to the minima in $k_{xx}$ at $H_{C1}$ and $H_{C2}$, increased phonon scattering by soft magnetic excitations can explain the minima at $H_{D1}$ and $H_{D2}$. However, whilst the anomalies in $k_{xx}$ and d$M$/d$H$ at the well-defined phase transitions at $H_{C1}$ and $H_{C2}$ remain sharp down to lowest temperature (Fig. 1b and SM), the anomalies at $H_{D1}$ and $H_{D2}$ are broader and weaker. They are enhanced upon cooling to 500 mK but subsequently suppressed below 500 mK, indicative of a reduction of soft magnetic excitations. These differences strongly suggest that $H_{D1}$ and $H_{D2}$ are not sharp and well-defined phase transitions like $H_{C1}$ and $H_{C2}$ but highly likely crossovers with soft but finite energy excitations or alternatively very weak transitions with only a limited number of low-energy excitations involved.

Recent reports of the magnetocaloric effect[23] and the magnetic Grüneisen parameter[25] at temperatures above 1 K also showed anomalies near $H_{D1}$ and $H_{D2}$, whilst the specific heat[24] at 0.67 K reported a symmetry-breaking phase transition near 10 T, a conclusion which is contested by several other measurements[23,25,27,28]. In the course of preparing this manuscript, we noticed the work by Czajka and coworkers[22], who also report a field-dependent structure in

$k_{XX}$ and ascribe it to quantum oscillations from underlying quasi-particles. That scenario explains neither the disappearance of features in the low temperature limit, nor why prominent minima align with the well-established magnetic phase transitions.

**Thermal Hall conductivity close to the half-quantized value**

A strong $k_{XY}/T$ signal is resolved above $H_{C2} = 7.1$ T, as measured by sweeping temperature at a fixed field (*T*-sweeps in Fig. 2) and magnetic field at a fixed temperature (*H*-sweeps in Fig. 3). Great care was taken to avoid systematic errors by verifying the linear power dependence of the signal and by avoiding field-hysteretic effects that may mix $k_{XX}$ and $k_{XY}$ (see: SM). Data from the *H*-sweeps and those from the *T*-sweeps are indeed consistent with each other within the given error bars, which arise from random noise in thermometry (see black points in Fig. 2 and thick lines in Fig. 3 for comparison).

As seen in Figs. 2 and 3, $k_{XY}/T$ is larger than the half-quantized value $k_{HQ}/T = 0.87$ mW/K²m at high temperatures and at high magnetic fields (above $T = 6.5$ K and $H_{C2} = 7.1$ T) and decreases to $k_{HQ}/T$ and below by lowering $T$ and $H_{\parallel}$, giving rise to a region with $k_{XY}/T \approx k_{HQ}/T$ on the $H_{\parallel}$-$T$ plane. A color plot of $k_{XY}/T$ across the whole phase diagram in Fig. 4 combines data from all the *H*- and the *T*-sweeps, which maps out the region where $k_{XY}/T$ is within ±20 % of $k_{HQ}/T$ in white. We see an *L*-shaped white region of $k_{XY}/T \approx k_{HQ}/T$, which runs first vertically from $T = 9$ K to $T \sim$ 6.5 K at around $H_{\parallel} = 7.6$ T and then continues horizontally from $H_{\parallel} \sim 7.6$ T up to at least $H_{\parallel} \sim$ 13.2 T at around 6.5 K. The observation of a $k_{XY}/T \approx k_{HQ}/T$ plateau as a function of $H_{\parallel}$ around $T$ = 6.5 K essentially reproduces those in previous studies, but its field range extends up to a higher field. The difference between reports might be ascribed to a sample-dependent strength of the effect of the features around 10 T on $k_{XY}/T$ (see: discussion below). The most intriguing feature we observed in Fig. 4 is the significant broadening of the white region to low temperatures with increasing field above ~10 T, giving rise to a large, triangular white area which extends down to at least 2 K at 13 T.

Does the observation of a region with $k_{XY}/T \approx k_{HQ}/T$ on the $H_\parallel$-$T$ plane have a physical significance and support the presence of the topologically protected half-quantized plateau? The enhancement of $k_{XY}/T$ could have different origins, for instance through (non-quantized) topological magnon transport[5,6]. In such scenarios, a crossing may give rise to an accidental $k_{XY}/T(H_\parallel,T) \approx k_{HQ}/T$-line on the $H_\parallel$-$T$ plane. In contrast, a quantized, topologically protected $k_{XY}/T(H_\parallel,T) \approx k_{HQ}/T$ plateau should exhibit insensitivity to changes both in magnetic field and temperature across and extended area in $H_\parallel$ and $T$.

The extended nature of $k_{XY}/T(H_\parallel,T) \approx k_{HQ}/T$ above 10 T represented by the white triangular plane in Fig. 4 argues against an accidental crossing scenario and in favor of a half-quantized plateau and hence a Majorana edge state in α-RuCl$_3$. The corresponding $T$-sweep data in Fig. 2 shows that, starting at 10.3 T, a kink-like singularity in the $T$-dependence of $k_{XY}/T$ appears at ~6.5 K (Fig. 2, red arrows). Notably, at the kink, $k_{XY}/T$ is always ~ $k_{HQ}/T$, implying that the magnitude of $k_{HQ}/T$ has special significance in the system. Upon lowering $T$, the kink is followed by an almost $T$-independent $k_{XY}/T \approx k_{HQ}/T$ region, reminiscent of an incipient half-quantized $T$-plateau, and then by a rapid decrease to zero. The width of the $T$-plateau-like region expands gradually upon increasing field from 10.3 T, giving rise to the white triangle in Fig. 4.

The narrow L-shaped white region below $H_\parallel \sim 10$ T in Fig. 4, on the other hand, represents a "line" of $k_{XY}/T(H_\parallel,T) \approx k_{HQ}/T$, which is clearly demonstrated by the $T$- and $H$- sweeps in Figs. 2 and 3. Along the vertical white region in Fig. 4, $k_{XY}/T$ is relatively insensitive to $T$ in the $T$-sweeps at 7.6 T (Fig. 2) but decreases continuously through $k_{HQ}/T$ with decreasing $H_{\parallel}$ in the $H$-sweeps (Fig. 3). Along the horizontal white region at around 6.5 K, $k_{XY}/T$ shows a plateau in the $H$-sweep at 6.5 K (Fig. 3) but decreases continuously through $k_{HQ}/T$ in the $T$ sweeps up to ~10T. It is not obvious why $k_{XY}/T$ crosses $k_{HQ}/T$ at a constant temperature of 6.5 K and a constant field of 7.6 T. Nevertheless, because of the line character of the region of $k_{XY}/T(H_{\parallel},T) \approx k_{HQ}/T$, we cannot rule out the scenario of an accidental crossing below $H_{\parallel}$ ~ 10 T.

**Low temperature suppression of $k_{XY}/T$**

What is the origin of rapid suppression of $k_{XY}/T$ to zero at low temperatures below 12 T, seen as the red colored region in Fig. 4? Within the red region, weak structures are observed: three vertical streaks of minima (deeper red) at around $H_{\parallel}$ = 7 T, 9.5 T and 11 T and two vertical streaks of maxima (brighter red) around $H_{\parallel}$ = 8.5 T and 10 T branching out from the white region at $T$ = 6.5 K. As seen in Figs. 1d and 4, the fields of the three minima apparently coincide with the critical field $H_{C2}$ and the two high field anomalies at $H_{D1}$ and $H_{D2}$ observed in $k_{XX}$ and $dM/dH$. The ~10 T maximum reflects the singularly weak $T$-dependence of $k_{XY}/T$ at $H_{\parallel}$ = 10.3 T (Fig. 2) and coincides with $H_P$. These close correlations strongly suggest that the high field magnetic crossovers/transitions are the cause of the suppression of $k_{XY}/T$ at low temperatures. It may be tempting to infer that, if there were no high filed anomalies at $H_{D1}$ and $H_{D2}$, the white region would cover a much wider region of the $H_{\parallel}$-$T$ plane, above $H_{C2}$ and below $T$ = 6.5 K.

If a half-quantized thermal Hall plateau arises from a chiral Majorana edge mode[8], it was shown theoretically that coupling between a phonon bath and the edge mode[30,31] is necessary to observe the plateau in the presence of dominant phonon conductivity. Phonons must be in the diffuse rather than the ballistic scattering regime. If phonons experience a crossover from the diffuse to the ballistic regimes upon cooling or under magnetic field, the thermal Hall signal from the Majorana edge mode should vanish due to the decoupling of phonons and edge modes, which might account for the low temperature suppression of $k_{XY}/T$. From the estimation of the phonon mean free path ($l_{ph}$) from $k_{XX}$ as a function of $H_\parallel$ and $T$, however, we may exclude the phonon-edge mode decoupling scenario as the origin of low temperature suppression of $k_{XY}/T$ below $k_{HQ}/T$ from 7 T to 12 T.

$k_{XX}$ approaches a $T^3$ power law at highest fields and lowest temperatures (Fig. 1f), consistent with phonon transport limited by a temperature-independent scattering length. The calculated $l_{ph}$ (see: SM) reaches ~50 µm at lowest temperatures, which is somewhat smaller than the sample width of ~1 mm (Fig. 1f), meaning that phonon transport at low temperatures is marginally in the diffuse limit. Furthermore, the fact that the regions of suppressed $k_{XY}/T$ coincide with minima in $k_{XX}$, i.e. regions of *shortest* $l_{ph}$, argues against the decoupling scenario. The low temperature enhancement of $k_{XY}/T$ to $k_{HQ}/T$ upon increasing magnetic field is accompanied by an *increase* of $l_{ph}$, the opposite of what would be expected from phonon-edge mode decoupling. We also note that the decoupling scenario is contrary to reports on the sample purity dependence of the half-quantized plateau in α-RuCl$_3$[4].

**Conclusion**

A comprehensive study of the thermal Hall effect across the phase diagram with in-plane field $H_\parallel$ // a reveals an extended region with $k_{XY}/T$ close to the half-quantized value, in particular below ~6.5 K and starting at ~10 T, well above the critical field $H_{C2}$ = 7.1 T. Alternative mechanisms leading to an enhancement $k_{XY}/T$ to $k_{HQ}/T$, such as a topological magnon-driven

Hall effect[5,6], have been proposed. Our observation of a robust $k_{XY}/T \approx k_{HQ}/T$ plane below the kink at $T = 6.5$ K, however, is in favor of the presence of a half-quantized plateau associated with a Majorana chiral edge state. Suppressions of $k_{XY}/T$ below the half-quantized value are observed below ~12 T at low temperature, which we argue to originate from high field anomalies representing crossovers or weak transitions identified in $k_{XX}$ and $dM/dH$. The identification of possible topological phase transitions into and out of the half-quantized state still remains an open question[23-25,27,28]. For the former, unveiling the nature of high field anomalies should be the key question to be addressed. Our data show that the latter must be at a field higher than 13 T. A measurement of $k_{XY}/T$ to much higher fields is highly desired.

**Methods**

**Samples.** Large, thin (~ 2.5 mm x 1.3 mm x 17 μm), high quality single crystals of α-RuCl$_3$ from the same growth batches as those reported in ref. 2 were measured. The sample we investigate for thermal Hall measurements has a confirmed half-quantized Hall plateau in that report, where it was labelled sample 2. A second sample with a similarly high thermal conductivity was used for magnetic susceptibility measurements.

**Thermal conductivity** was measured using a steady-state three-thermometer setup. For the temperature range 1.8 K - 9 K Cernox-1050 chip thermometers (Lakeshore) were used in a $^4$He cryostat. For the temperature range 150 mK – 3 K ruthenium oxide chip thermometers were used in a dilution fridge. In each case, thermometers were calibrated in situ against a field-calibrated reference thermometer. A 10 kΩ NiCr resistive heater was used to apply heater power. The sample was mounted with Apiezon N grease onto a LiF single crystal, which was attached with silver paint onto a copper mount, which was cooled by the cryostat. Several copper mounts machined at different angles were used to allow for measurements in fixed tilted fields.

We adopt the usual convention that the crystal *a*-axis is perpendicular to Ru-Ru bonds, the *b*-axis lies along the Ru-Ru bonds, and the *c*-axis is perpendicular to the honeycomb plane. Magnetic field was applied at either 70 degrees from c-axis ($^4$He), or 90 degrees from c-axis (dilution refrigerator), with the in-plane field component in the a-axis. The amplitude of $k_{XX}$, as well as the measured transition fields ($H_{C1}$, $H_{C2}$, $H_{D1}$, $H_{D2}$), were confirmed to reproduce at the same values of $H_\parallel$ for these two field angles, so the two data sets are presented as one. The field angle was determined under an optical microscope with 1° precision. Heat was always applied along the *a*-axis.

Over the course of these measurements, the crystal was thermally cycled between room temperature and < 4.2 K 14 times, and each time identical values of $k_{XX}$ were measured, indicating that sample deterioration over time is negligible. The measurement of $k_{XX}$ is additionally in very close agreement with that previously measured in Kyoto (SM, Fig. S1). Experimentally, the existence of a half-quantized thermal Hall plateau has been discussed to correlate with the magnitude of the low temperature thermal conductivity[4]. We measure a conductivity of 7 W/Km at 5 K (Fig. 1e), which exceeds that of the cleanest sample in that report.

**Magnetization** measurements above 2K were performed using a PPMS VSM option (Quantum Design), and below 1K using a custom-built Faraday force magnetometer on a dilution refrigerator. For all magnetization data the applied field was parallel to the crystal *a*-axis.

# Figures

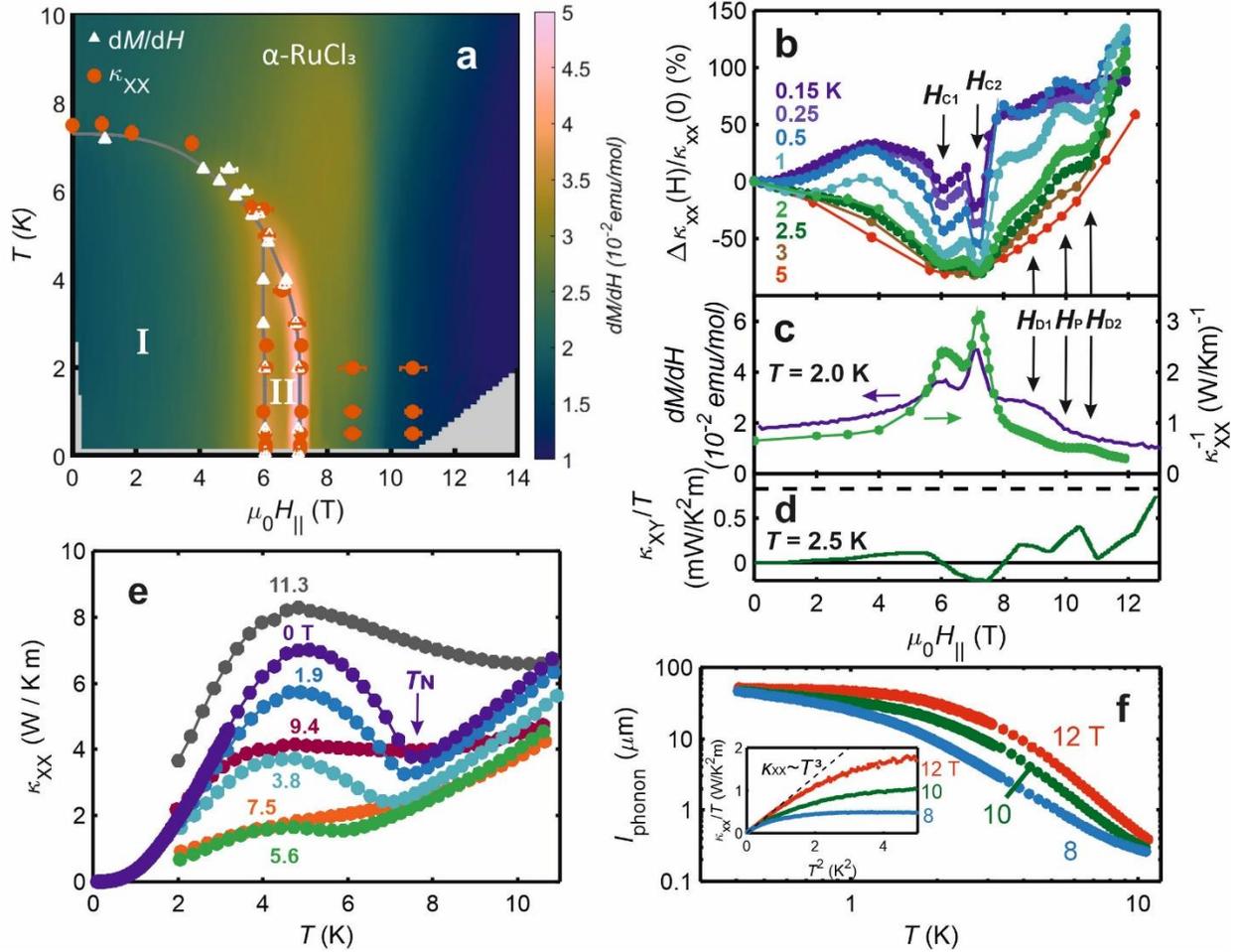

**Fig.1 Anomalies of thermal conductivity $k_{xx}$ in magnetic field and phase diagram. a**, $H_{\parallel}, T$ phase diagram of α-RuCl$_3$ derived from features in the thermal conductivity ($k_{XX}$) and magnetic susceptibility (d$M$/d$H$). Phases I and II were previously identified as regions of long-range antiferromagnetic order (enclosed by solid grey lines). Color plot: interpolated magnitude of d$M$/d$H$ (Raw data in SM Fig. S2). White triangles mark local maxima in d$M$/d$H$ and d$M$/d$T$, orange points mark minima in $k_{XX}$. **b**, Isotherms of the relative change of $k_{XX}$ with respect to the zero field value, as a function of magnetic field. Antiferromagnetic transitions $H_{C1}$, $H_{C2}$, as well as features at $H_{D1}$, $H_{D2}$ appear as local minima of $k_{XX}$, the feature at $H_P$ appears as a broad local maximum. **c**, Comparison of the sequence of anomalies in d$M$/d$H$ and $k_{XX}^{-1}$ as a function of magnetic field at 2.0 K. **d**, Thermal Hall conductivity divided by temperature ($k_{XY}/T$) as a function of magnetic field at 2.5K. The data are interpolated from the data set shown in Fig 4, and the dashed line is the half-quantized value $k_{HQ}/T$. **e**, $k_{XX}$ as a function of temperature measured in a $^4$He flow cryostat, for in-plane field values from 0 to 11.3 T. For 0 T, a low-temperature dilution refrigerator measurement is superimposed. The antiferromagnetic transition temperature $T_N$ is identified as a sharp minimum. **f**, Insert: $k_{XX}/T$ at high fields, demonstrating a low-temperature tendency to phonon-driven $k_{XX} \sim T^3$ behavior (dashed guide to the eye). Main figure: calculated phonon mean free path as a function of temperature. At 12 T, the lowest temperature phonon mean free path saturates around 50 μm.

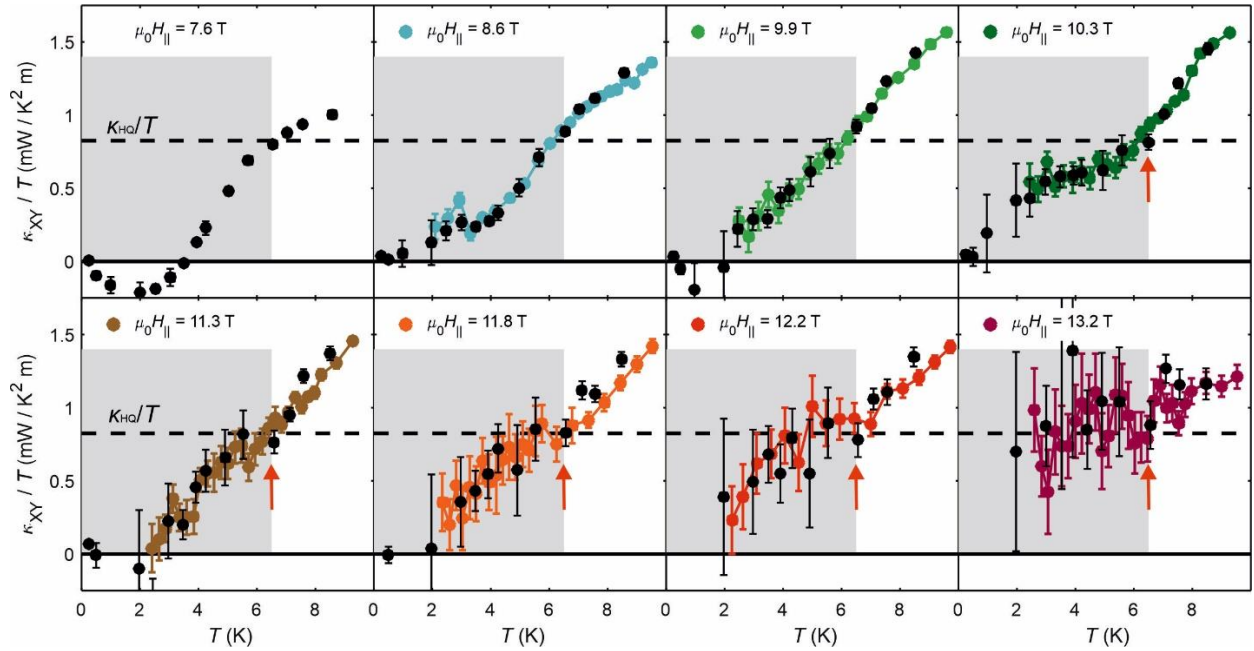

**Fig. 2 Temperature dependence of thermal Hall effect $k_{XY}/T$.** The panels display temperature-dependent $k_{XY}/T$ data in order of increasing in-plane magnetic field from 7.6 T (top left) to 13.2 T (bottom right). The independent results of temperature sweeps at fixed field (colored points) and field sweeps at fixed temperature (black points) are plotted together, to demonstrate the high degree of consistency. From 10.3 T onward, a kink is visible around 6.5 K (red arrows), and beyond this field the low temperature magnitude of $k_{XY}/T$ is gradually enhanced towards the half-quantized value $k_{HQ}/T$ (dashed line). The grey shaded areas emphasize the temperature range below 6.5 K.

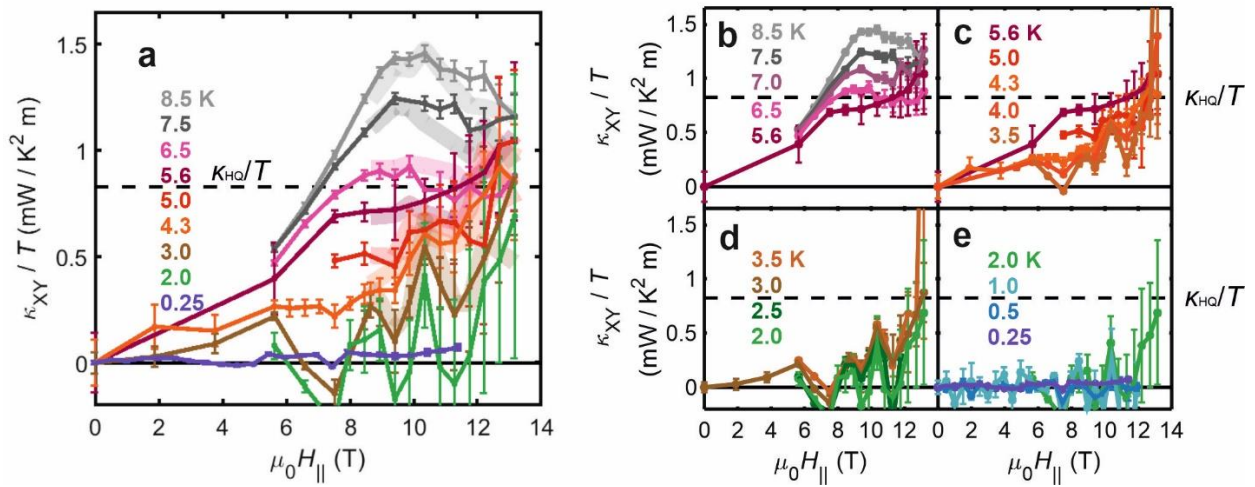

**Fig. 3 Magnetic field dependence of thermal Hall effect $k_{XY}/T$. a**, Selection of $k_{XY}/T$ isotherms as a function of in-plane magnetic field from 250 mK to 8.5 K. A strong and complex dependence on $T$ and $H_\parallel$ is observed, including enhancement at highest fields towards the half-quantized value $k_{HQ}/T$ (dashed line), suppression upon cooling at lower fields, and non-monotonic structure for 0.25 K < $T$ < 4 K. Broad, shaded lines indicate trends interpolated temperature sweeps at fixed field (Fig. 2), to demonstrate consistency. **b-e**, extended selection of $k_{XY}/T$ isotherms in order of decreasing temperature.

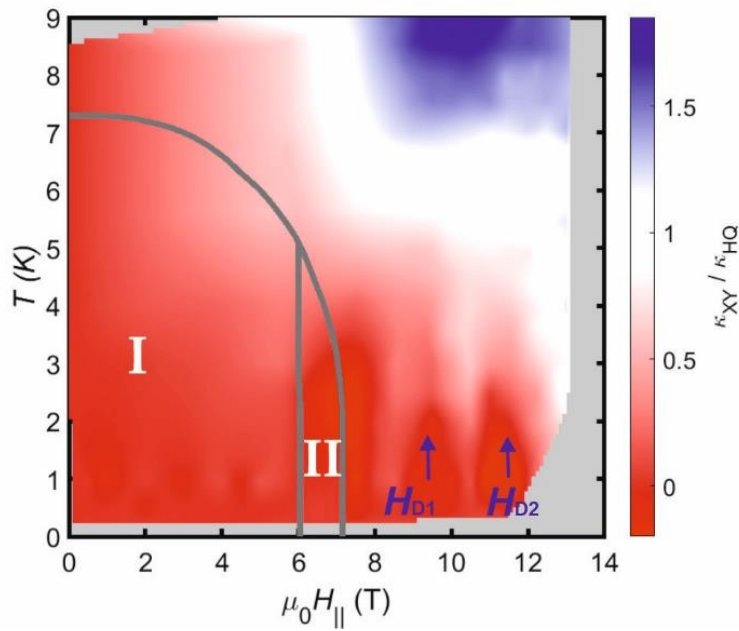

**Fig. 4 Overview of thermal Hall effect $k_{XY}/T$ on the $H_\parallel$-$T$ plane.** Interpolated magnitude of $k_{XY}/T$ normalized by $k_{HQ}/T$ across the $H_\parallel$,$T$ phase diagram. In the color scheme, white represents a thermal Hall conductivity within 20% of the half-quantized value. The grey lines trace the antiferromagnetic phase transitions, for reference. The two low temperature minima in $k_{XY}/T$ at fields $H_{D1}$ and $H_{D2}$ are indicated with arrows.

**Acknowledgements**

We thank Y. Kasahara and Y. Matsuda for insightful discussions and M. Dueller and K. Pflaum for technical assistance. The work done in Germany has been supported in part by the Alexander von Humboldt foundation. H.TAN and N.K. have been supported by JSPS KAKENHI Grant Number JP17H01142 and JP19K03711, respectively. H.TAK has been supported in part by JSPS KAKENHI, Grant Number JP17H01140.

**Author Contributions**

J.A.N.B, R.R.C, Y.M and H.TAK conceived the research. N.K and H.TAN synthesized the single crystals. J.A.N.B and R.R.C designed and performed thermal conductivity experiments. Y.M. designed and performed the magnetic susceptibility experiments. J.A.N.B, R.R.C, Y.M and H.TAK analyzed data and participated in the writing of the paper. All authors contributed to the manuscript preparation.

**Competing interests**

The authors declare no competing interests.

**References**

1. Balents, L. Spin liquids in frustrated magnets. *Nature* **464**, 199–208 (2010).

2. Kasahara, Y. *et al.* Majorana quantization and half-integer thermal quantum Hall effect in a Kitaev spin liquid. *Nature* **559**, 227–231 (2018).

3. Yokoi, T. *et al.* Half-integer quantized anomalous thermal Hall effect in the Kitaev material α-RuCl$_3$. Preprint at http://arxiv.org/abs/2001.01899v1 (2020).

4. Yamashita, M., Kurita, N. & Tanaka, H. Sample dependence of the half-integer quantized thermal Hall effect in a Kitaev candidate α-RuCl$_3$. *Phys. Rev. B* **102**, 220404 (2020).

5. McClarty, P. A. et al. Topological magnons in Kitaev magnets at high fields. Phys. Rev. B **98**, 060404 (2018).

6. Zhang, E. Z., Chern, L. E. & Kim, Y. B. Topological magnons for thermal Hall transport in frustrated magnets with bond-dependent interactions. Preprint at http://arxiv.org/abs/2102.00014v1 (2021).

7. Kitaev, A. Anyons in an exactly solved model and beyond. *Ann. Phys.* **321**, 2–111 (2006).

8. Nasu, J., Yoshitake, J. & Motome, Y. Thermal Transport in the Kitaev Model. *Phys. Rev. Lett.* **119**, 127204 (2017).

9. Motome, Y. & Nasu, J. Hunting Majorana fermions in kitaev magnets. *J. Phys. Soc. Japan* **89**, 012002 (2020).

10. Jackeli, G. & Khaliullin, G. Mott insulators in the strong spin-orbit coupling Limit: From Heisenberg to a Quantum Compass and Kitaev Models. *Phys. Rev. Lett.* **102**, 017205 (2009).

11. Trebst, S. Kitaev Materials. Preprint at http://arxiv.org/abs/1701.07056v1 (2017).

12. Winter, S. M., Li, Y., Jeschke, H. O. & Valentí, R. Challenges in design of Kitaev materials: Magnetic interactions from competing energy scales. *Phys. Rev. B* **93**, 214431 (2016).

13. Takagi, H., Takayama, T., Jackeli, G., Khaliullin, G. & Nagler, S. E. Concept and realization of Kitaev quantum spin liquids. *Nat. Rev. Phys.* **1**, 264–280 (2019).

14. Baek, S. H. *et al.* Evidence for a field-induced quantum spin liquid in α-RuCl$_3$. *Phys. Rev. Lett.* **119**, 037201 (2017).

15. Wang, Z. *et al.* Magnetic Excitations and Continuum of a Possibly Field-Induced Quantum Spin Liquid in α-RuCl$_3$. *Phys. Rev. Lett.* **119**, 227202 (2017).

16. Zheng, J. *et al.* Gapless Spin Excitations in the Field-Induced Quantum Spin Liquid Phase of α-RuCl$_3$. *Phys. Rev. Lett.* **119**, 227208 (2017).

17. Banerjee, A. *et al.* Excitations in the field-induced quantum spin liquid state of α-RuCl$_3$. *npj Quantum Mater.* **3**, 8 (2018).

18. Johnson, R. D. *et al.* Monoclinic crystal structure of α-RuCl$_3$ and the zigzag antiferromagnetic ground state. *Phys. Rev.* **92**, 235119 (2015).

19. Kubota, Y., Tanaka, H., Ono, T., Narumi, Y. & Kindo, K. Successive magnetic phase transitions in α-RuCl$_3$: XY-like frustrated magnet on the honeycomb lattice. *Phys. Rev. B.* **91**, 094422 (2015).

20. Yadav, R. *et al.* Kitaev exchange and field-induced quantum spin-liquid states in honeycomb α-RuCl$_3$. *Sci. Rep.* **6**, 37925 (2016).

21. Sears, J. A. *et al.* Magnetic order in α-RuCl$_3$: A honeycomb-lattice quantum magnet with strong spin-orbit coupling. *Phys. Rev. B* **91**, 144420 (2015).

22. Czajka, P. *et al.* Oscillations of the thermal conductivity observed in the spin-liquid state of α-RuCl$_3$. Preprint at http://arxiv.org/abs/2102.11410v1 (2021).

23. Balz, C. *et al.* Finite field regime for a quantum spin liquid in α-RuCl$_3$. *Phys. Rev. B* **100**, 60405 (2019).

24. Tanaka, O. *et al.* Thermodynamic evidence for field-angle dependent Majorana gap in a Kitaev spin liquid. Preprint at http://arxiv.org/abs/2007.06757v1 (2020).

25. Bachus, S. *et al.* Thermodynamic Perspective on Field-Induced Behavior of α-RuCl$_3$. *Phys. Rev. Lett.* **125**, 97203 (2020).

26. Lampen-Kelley, P. et al. Field-induced intermediate phase in α-RuCl$_3$: Non-coplanar order, phase diagram, and proximate spin liquid. Preprint at http://arxiv.org/abs/1807.06192v1 (2018).

27. Ponomaryov, A. N. *et al.* Nature of Magnetic Excitations in the High-Field Phase of α-RuCl$_3$. *Phys. Rev. Lett.* **125**, 37202 (2020).

28. Schönemann, R. *et al.* Thermal and magnetoelastic properties of α-RuCl$_3$ in the field-induced low-temperature states. *Phys. Rev. B* **102**, 214432 (2020).

29. Hentrich, R. *et al.* Unusual Phonon Heat Transport in α-RuCl$_3$: Strong Spin-Phonon Scattering and Field-Induced Spin Gap. *Phys. Rev. Lett.* **120**, 117204 (2018).

30. Vinkler-Aviv, Y. & Rosch, A. Approximately Quantized Thermal Hall Effect of Chiral Liquids Coupled to Phonons. *Phys. Rev. X* **8**, 31032 (2018).

31. Ye, M., Halász, G. B., Savary, L. & Balents, L. Quantization of the Thermal Hall Conductivity at Small Hall Angles. *Phys. Rev. Lett.* **121**, 147201 (2018).

# Supplementary Materials

## Absolute signal reproducibility in $k_{XX}$: Kyoto and Stuttgart

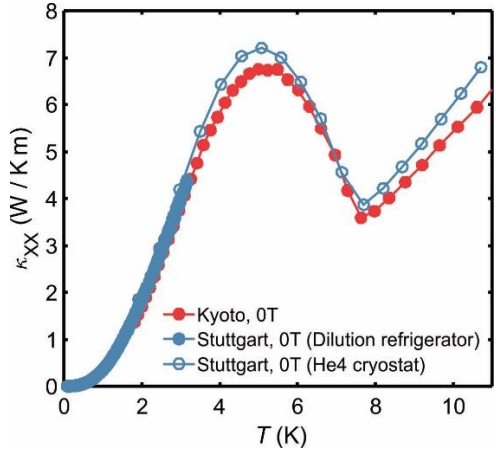

**Fig. S1** Comparison of three independent measurements of the thermal conductivity ($k_{XX}$) of sample 2. Red circles: data measured in Kyoto, Solid blue circles: data taken in a dilution refrigerator in Stuttgart, open blue circles: data taken in a $^4$He flow cryotstat in Stuttgart. Each measurement has slightly modified contact placements.

## Magnetization isotherms and phase transitions

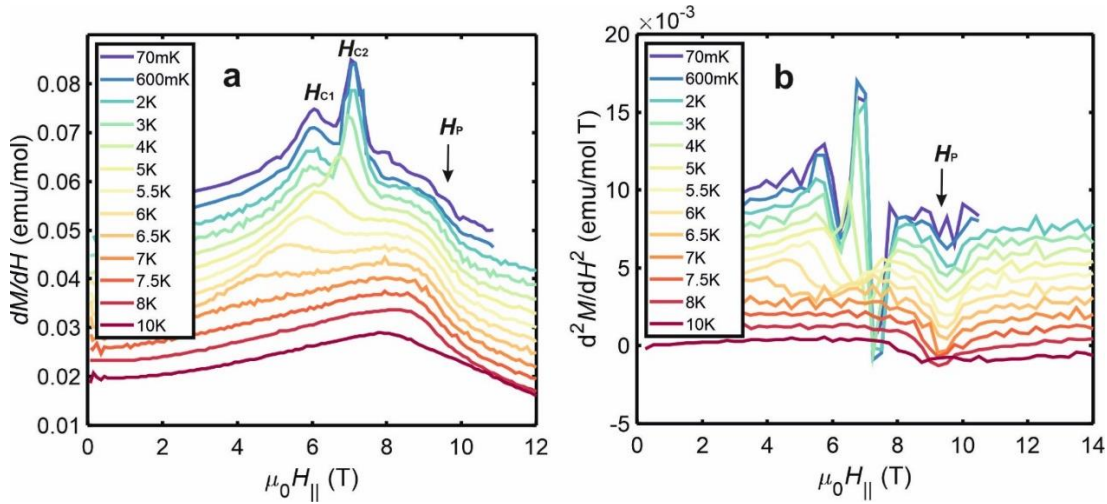

**Fig. S2** Field dependences of d$M$/d$H$ which were used to construct the phase diagram in Fig. 1a (main text). **a**, isotherms of d$M$/d$H$, with every curve offset by 0.003 emu/mol for clarity. Phase transitions out of the antiferromagnetic phases $H_{C1}$, $H_{C2}$ are identified by peaks in d$M$/d$H$, whereas $H_P$ coincides with a drop in d$M$/d$H$. **b**, isotherms of d$^2M$/d$H^2$, where the feature at $H_P$ is identified as a local minimum. $H_P$ shifts only minimally in field upon heating, and is no longer resolved at 10 K. Curves are offset by 8x10$^{-4}$ emu/mol T for clarity.

## Calculation of $k_{XX}$, $k_{XY}$

The thermal conductivity tensor $k$ of a material is given by $j = -\kappa \nabla T$ where $j = P/A$ is the heat flux arising from power $P$ and the material cross section $A = w \times t$ where $w$ ($t$) is the sample width (thickness), taken to be constant along the direction of heat flow in the case of a bar-shaped shape. In an experiment three thermometer contacts are placed on the sample measuring the longitudinal temperature difference $\Delta T_X$ along length $L$ parallel to the heat current and the transverse temperature difference $\Delta T_Y$ along width $w_H$ perpendicular to the heat current. $w_H$ may be slightly smaller than $w$ when partially cover the top of a sample. Assuming isotropic heat flow in the 2D plane and in the presence of an applied magnetic field $H$ we obtain

$$\frac{1}{wt}\begin{pmatrix} P \\ 0 \end{pmatrix} = \begin{pmatrix} \kappa_{xx}(H) & \kappa_{xy}(H) \\ -\kappa_{xy}(H) & \kappa_{xx}(H) \end{pmatrix} \cdot \begin{pmatrix} \Delta T_x(H)/L \\ \Delta T_y(H)/w_H \end{pmatrix}$$

To avoid a longitudinal response in $\Delta T_Y$ owing to possible contact misalignment, the signal is antisymmetrized by measuring with applied positive and negative magnetic field $\pm H$:

$$\Delta T_y^{asym}(H) = [\Delta T_y(+H) - \Delta T_y(-H)]/2$$

$$\Delta T_x^{sym}(H) = [\Delta T_x(+H) + \Delta T_x(-H)]/2$$

The thermal conductivities are then calculated as

$$\kappa_{xx} = \frac{P \cdot \Delta T_x^{sym}}{L \cdot w \cdot t \cdot \left(\left(\frac{\Delta T_x^{sym}}{L}\right)^2 + \left(\frac{\Delta T_y^{asym}}{w_H}\right)^2\right)}$$

$$\kappa_{xy} = \frac{P \cdot \Delta T_y^{asym}}{w_H \cdot w \cdot t \cdot \left(\left(\frac{\Delta T_x^{sym}}{L}\right)^2 + \left(\frac{\Delta T_y^{asym}}{w_H}\right)^2\right)}$$

Here power $P = I^2 \times R$ with heater with resistance $R$ and applied electrical current $I$. In our setup we calculate the heat leakage (heater power bypassing the sample by conduction through wires and mounts to the cold finger) to be at least two orders of magnitude smaller than the heat flow through the sample at all temperatures.

**Field-hysteresis induced mixing of $k_{XX}$, $k_{XY}$**

In an ideal measurement, the transverse temperature response between 'Hall' and 'cold' thermometers $\Delta T_Y = T_{hall} - T_{cold}$ is perfectly asymmetric and has no longitudinal component, but in real samples there is always some longitudinal component due to contact misalignment. The measured longitudinal signal $\Delta T_X = T_{hot} - T_{cold}$ arises over distance L (contact separation). Distance L* defines the contact offset between $T_{hall}$ and $T_{cold}$ along the direction of heat flow, which can be either positive or negative.

If the purely intrinsic Hall signal is $\Delta T_{Yi}$, then $\Delta T_Y = \Delta T_{Yi} - (L^*/L)\Delta T_X$

We calculate the Hall temperature difference by field-antisymmetrisation as above, so only a field-asymmetric component of the measured longitudinal gradient $\Delta T_X$ couples into $\Delta T_Y^{asym}$. Such an asymmetric component may arise from hysteresis in field, which may be caused by magnet hysteresis or sample-intrinsic hysteresis. In particular, the error is expected to increase proportionally to a) the contact offset: L*, b) the field derivative of $k_{XX}$: $dk_{XX}/dH$ and c) the magnitude of field hysteresis.

To ensure symmetry in any hysteretic components, only 'inward' (absolute field decreasing) or 'outward' field sweeps (absolute field increasing) are compared for analysis. Furthermore, we can compare the measurements for 'inward' sweeps with two distinct contact offset configurations: one where L* is positive, and one where L* is negative (Fig. S3). In both cases, the features in $k_{XY}/T$ have the same sign. If the features were caused by hysteresis and mixing of $k_{XX}$ into $k_{XY}$, the two configurations should have had opposite sign. We therefore conclude that the minima in the field dependence of $k_{XY}/T$ are intrinsic. Furthermore, the comparison between the wider field and temperature dependence of $k_{XY}/T$ for configuration 2 (main text, Fig. 3) and configuration 1 (Fig. S3e) reveals that all main features in the data are reproduced.

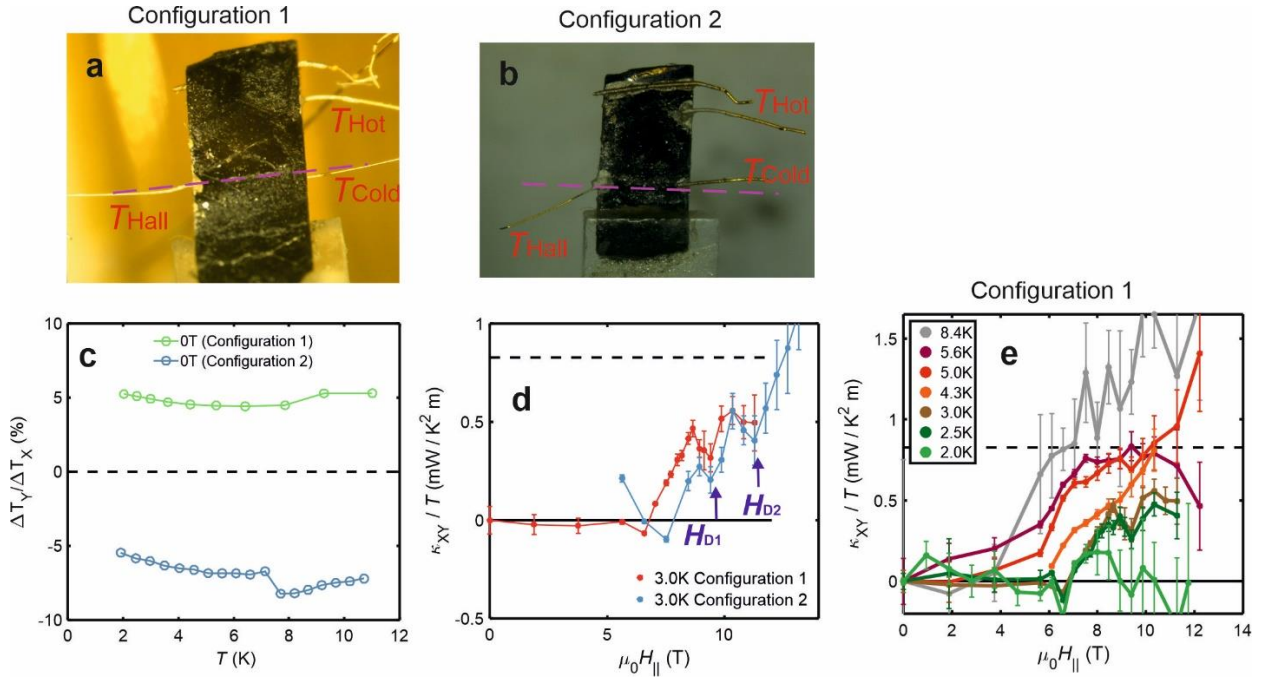

**Fig. S3 a,b,** photographs of sample 2 with wires attached in configurations 1 and 2. Purple dashed line are guides to the eye to help identify the contact offset. **c,** measured Hall contact offset expressed as $\Delta T_Y / \Delta T_X$ (%) at 0 T, showing the expected inverse offsets for the two configurations. In the absence of an offset, $\Delta T_Y / \Delta T_X$ would be zero (dashed line). **d,** isotherms of $k_{XY}/T$ at 3.0K for both contact configurations, showing the same features at $H_{D1}$ and $H_{D2}$. **e,** isotherms of $k_{XY}/T$ for configuration 1 at multiple temperatures, demonstrating the reproducibility of the main features of configuration 2 (main text, Fig. 3).

**Excluding power dependence in $k_{XY}$**

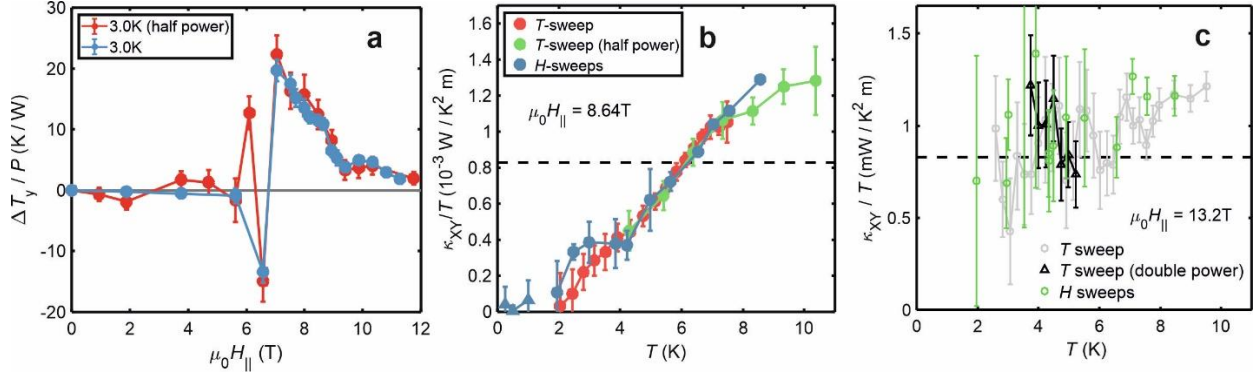

**Fig. S4** Repeated measurements at different levels of heater power to exclude the possibility of systematic power dependent error. **a,** Isotherms of $\Delta T_Y/P$ at 3.0 K. **b,** temperature dependences of $k_{XY}/T$ at 8.64 T. **c,** temperature dependences of $k_{XY}/T$ at 13.2 T.

To test the validity of thermal conductivity data, it is necessary to verify whether the measurement is ohmic, i.e. that the measured temperature gradients in the steady state are linearly proportional to the applied heater power, and that the reported thermal conductivity and thermal Hall conductivity are therefore independent of the heater power. Here, we present several test measurements which demonstrate an ohmic response in the thermal Hall data. For the data presented in the main text, the temperature gradient along the sample $\Delta T_X$ was maintained around 10% of absolute temperature. In Fig. S4a, the Hall temperature gradient normalized by heater power ($\Delta T_Y/P$) is shown for two independent isotherms at 3K, one of which was performed at half the standard heater power (red markers, $\Delta T_X/T \approx 5\%$). The two curves overlap as expected. In Fig. S4b, the temperature dependence of $k_{XY}/T$ at 8.64T is compared for two $T$ sweeps, one of which was performed at half power (green markers, $\Delta T_X/T \approx 5\%$), together with full-power $H$ sweeps. All curves overlap within error. In Fig. S4c the temperature dependence of $k_{XY}/T$ is compared for two $T$ sweeps, one of which was performed at double the usual power ($\Delta T_X/T \approx 20\%$), together with normal-power $H$ sweeps. Again, the data overlap within the (substantially field-enhanced) error bars.

### Calculation of the phonon mean free path

Starting from the assumption that that phonon conductivity is approximately isotropic, the phonon mean free path may be calculated as $l_{ph} = 3 k_{XX} / C v$, where C is the phonon specific heat per unit volume and $v$ is the average phonon velocity. In the literature, slightly different low temperature phonon specific heat values are reported, with $C_{ph}/T^3 = 1.22$ mJ mol$^{-1}$K$^{-4}$ (ref. 1) and $C_{ph}/T^3 \approx 1.3$ mJ mol$^{-1}$K$^{-4}$ (ref. 2). The difference does not lead to significantly different mean free paths, and we perform the calculation using the former value.

We estimate the phonon velocity with the Debye relation $v = \left(\frac{2\pi^2 k_B^4}{5\beta\hbar^3}\right)^{\frac{1}{3}} \approx 1700 m/s$ where β = $C_{ph}/T^3$ as before. An overview of the calculated phonon mean free path across the phase diagram is presented in Fig. S5.

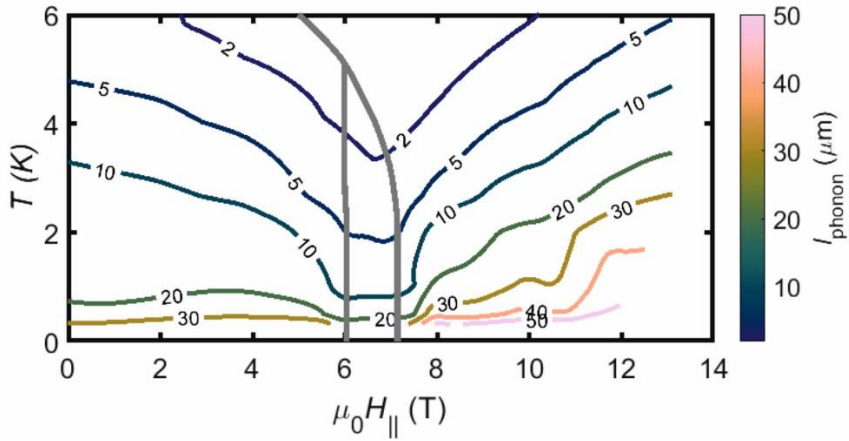

**Fig. S5** Contour plot of the calculated phonon mean free path ($l_{ph}$) interpolated across the $H_\parallel, T$ phase diagram. The grey lines trace the antiferromagnetic phase transitions, for reference.

### Field dependence of the thermal conductivity above 2.0 K

As shown in Fig S6, the sharp features in the field dependence of $k_{xx}$ that were detected at low temperature rapidly disappear upon heating above 3.0 K. At high temperatures, the field dependence displays a single, broad minimum around 7 T. At 8.6 K, which is above the ordering temperature $T_N$ = 7.5 K, the field dependence is dramatically weakened, although a high field upturn persists at this temperature.

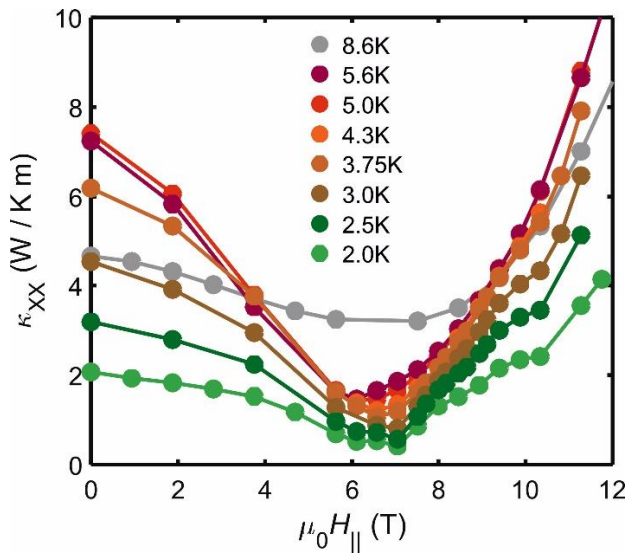

**Fig. S6)** Field dependence of $k_{xx}$ up to 12 T for the temperature range 2.0 K – 8.6 K.

**References**

1	Tanaka, O. *et al.* Thermodynamic evidence for field-angle dependent Majorana gap in a Kitaev spin liquid. Preprint at http://arxiv.org/abs/2007.06757v1 (2020).

2.	Widmann, S. et al. Thermodynamic evidence of fractionalized excitations in α-RuCl$_3$. Phys. Rev. B **99**, 094415 (2019).